\documentclass{pasa}%
\usepackage{epstopdf}       % to convert eps images to pdf images
\usepackage{verbatim}       % to allow multiline comments
\usepackage{bm}		  % allow shorthand to bold text
\usepackage{aas_macros}  % to allow journal shortcuts in the bibliography

\newcommand{\nicil}{{\sc Nicil}}
\newcommand{\nicilf}{{\tt nicil.F90}}
\newcommand{\nicilsource}{{\tt nicil\_source.F90}}
\newcommand{\nicileta}{{\tt nicil\_ex\_eta.F90}}
\newcommand{\niciletasup}{{\tt nicil\_ex\_eta\_sup.F90}}
\newcommand{\nicilsph}{{\tt nicil\_ex\_sph.F90}}
\newcommand{\nicilsphsup}{{\tt nicil\_ex\_sph\_sup.F90}}

\title[NICIL]{NICIL: A stand alone library to self-consistently calculate non-ideal magnetohydrodynamic coefficients in molecular cloud cores}
\author[James Wurster]{James Wurster$^{1,2}$\thanks{j.wurster@exeter.ac.uk}\\
\affil{$^1$School of Physics, University of Exeter, Stocker Rd, Exeter EX4 4QL, UK}
\affil{$^2$Monash Centre for Astrophysics and School of Physics and Astronomy, Monash University, Vic 3800, Australia}}
\jid{PASA}
\doi{10.1017/pas.\the\year.xxx}
\jyear{\the\year}

\usepackage[authoryear]{natbib}

\begin{document}%
\begin{abstract}
In this paper, we introduce \nicil: Non-Ideal magnetohydrodynamics Coefficients and Ionisation Library.  \nicil \ is a stand-alone {\sc Fortran90} module that calculates the ionisation values and the coefficients of the non-ideal magnetohydrodynamics terms of Ohmic resistivity, the Hall effect, and ambipolar diffusion.  The module is fully parameterised such that the user can decide which processes to include and decide upon the values of the free parameters, making this a versatile and customisable code.  The module includes both cosmic ray and thermal ionisation; the former includes two ion species and three species of dust grains (positively charged, negatively charged and neutral), and the latter includes five elements which can be doubly ionised.  We demonstrate tests of the module, and then describe how to implement it into an existing numerical code.
\end{abstract}
\maketitle
%------------------------------------------------------------------------------------------------------------------------------------------------------------------------------------------------------------------------------------------------------------------------------------------------------------
\section{INTRODUCTION }
\label{sec:intro}

Astrophysical magnetic fields are often modelled using ideal magnetohydrodynamics (MHD), which assumes that the gas is fully ionised and that the electrons are tied to the magnetic field and have zero resistivity.  However, in molecular cloud cores for example, this has been known to be a poor approximation to the true conditions since at least \citet{MestelSpitzer1956}, where the  ionisation levels can be as low as $n_\text{e}/ n_{\text{H}_2} = 10^{-14}$ (\citealt{NakanoUmebayashi1986,UmebayashiNakano1990}).    

When a medium is only partially ionised, such as in a molecular cloud, three non-ideal MHD effects occur: 
\begin{enumerate}
\item \emph{Ohmic resistivity}: the drift between electrons and ions/neutrals; neither ions nor electrons are tied to the magnetic field,
\item \emph{Hall effect}: ion-electron drift; only electrons are tied to the magnetic field, and 
\item \emph{Ambipolar diffusion}: ion-neutral drift; both ions and electrons are tied to the magnetic field. 
\end{enumerate}

These environmentally dependent terms add resistivity to the electrons, and ultimately affect the evolution of the magnetic field (e.g. \citealp{WardleNg1999}; \citealp{NakanoNishiUmebayashi2002}; \citealp{TassisMouschovias2007b}; \citealp{Wardle2007}; \citealp{PandeyWardle2008}; \citealp{KeithWardle2014}).  Recent studies, both idealised \citep[e.g.][]{Bai2014,Bai2015} and realistic (e.g. \citealp{ShuGalliLizanoCai2006,MellonLi2009,DuffinPudritz2009,DappBasu2010,KrasnopolskyLiShang2010,MachidaInutsukaMatsumoto2011,LiKrasnopolskyShang2011,DappBasuKunz2012,TomidaEtAl2013,TomidaOkuzumiMachida2015,TsukamotoEtAl2015b,TsukamotoEtAl2015,WPB2016}) have included some or all of the non-ideal MHD effects.  In each study, however, the authors have been required to write and implement their own algorithms to model the non-ideal MHD effects. 

Multiple algorithms existing within a community provides a self-consistent check on the processes, and this helps to validate the results.  However, writing and implementing the non-ideal MHD algorithms is a non-trivial process, both physically and numerically.  Thus, we introduce introduce \nicil: Non-Ideal MHD Coefficients and Ionisation Library, which is a pre-written, {\sc Fortran90} module that self-consistently calculates the non-ideal MHD coefficients and is ready for download and implementation into existing codes.  Early version of \nicil \ are used in \citet{WPB2016} and Wurster, Price \& Bate (submitted).  We are aware that a similar stand-alone code has recently been published by \citet{MarchandEtAl2016}.

This paper is a user-manual for \nicil, and is organised as follows.  Section~\ref{sec:algo} describes the algorithms that \nicil \ uses.  We perform a simple test of \nicil \ in Section~\ref{sec:tests}, and provide an astrophysical example in Section~\ref{sec:test:mc}.  Section~\ref{sec:imp} explains how to implement \nicil \ into an existing numerical code, and we conclude in Section~\ref{sec:conc}.  Throughout the text, we will refer to the ``author'' as the author of \nicil, the ``user'' as the person who is implementing \nicil \ into the existing code, and the ``user's code'' or ``existing code'' as the code into which \nicil \ is being implemented.  

%------------------------------------------------------------------------------------------------------------------------------------------------------------------------------------------------------------------------------------------------------------------------------------------------------------
\section{ALGORITHMS}
\label{sec:algo}

The complete \nicil \ library can be downloaded at www.bitbucket.org/jameswurster/nicil.  This is a free library under the {\tt GNU} license agreement: free to use, modify and share, does not come with a warranty, and this paper must be cited if \nicil \ or any modified version thereof is used in a study.   The important files are listed and summarised in Appendix~\ref{app:list}.  The important parameters which the user can modify, along with their default values, are listed in Appendix~\ref{app:fp}.

\nicil \ includes various processes that can be included or excluded at the user's discretion.  Major processes are cosmic ray and thermal ionisation (one or both can be selected) and the assumption of the grain size distribution (fixed or MRN distribution; \citealp*{MathisRumplNordsieck1977}).  By default, all three non-ideal MHD coefficients are self-consistently calculated, however, \nicil \ includes the option of calculating only selected terms or using fixed coefficients (see Appendix~\ref{app:fixed} for a discussion of fixed coefficients).  
 
The remainder of this section describes the processes that are used to calculate the non-ideal MHD coefficients.  We define all variables, but for clarity, we do not state the default values in the main text.  
%-----------------------------------------------------------------------------------------
\subsection{Cosmic ray ionisation}
\label{sec:algo:IR}
When a cosmic ray (or an X-ray) strikes an atom or molecule, it has the ability to remove an electron which creates an ion.  Two ion species are modelled: species i with mass $m_{\text{i}_\text{R}}$, which represents the light elements (e.g. hydrogen and helium compounds), and species I with mass $m_{\text{I}_\text{R}}$, which represents the metals.  The light ion mass is calculated from the given abundances or mass fractions of hydrogen and helium assuming all the hydrogen is molecular hydrogen, and the metals are modelled as a single species, where the mass is a free parameter.

Dust grains have the ability to absorb electrons, which gives the grains a negative charge or lose an electron to give them a positive charge.  \nicil \ includes two dust models: fixed radius and MRN size distribution.  For a fixed grain size, $a_\text{g}$, the number density is proportional to the total number density, $n$ \citep{KeithWardle2014}, 
\begin{equation}
\label{eq:ngrain}
n_\text{g} = \frac{m_\text{n}}{m_\text{g}}f_\text{dg} n,
\end{equation}
where $f_\text{dg}$ is the dust-to-gas mass ratio and $m_\text{n}$ is the mass of a neutral particle.  

The number density for the MRN grain distribution follows a power law, viz.,
\begin{equation}
\label{eq:mrnn}
\frac{\text{d}n_\text{g}}{\text{d}a} = An_\text{H}a^{-3.5},
\end{equation}
where $n_\text{H}$ is the number density of the hydrogen nucleus, $n_\text{g}(a)$ is the number density of grains with a radius smaller than $a$, and $A = 1.5\times 10^{-25}$ cm$^{2.5}$ \citep{DraineLee1984}.  For simplicity, we assume $n_\text{H}\approx n$, where $n$ is the total number density.   Following \citet{NakanoNishiUmebayashi2002}, the given range of $a_\text{n} < a < a_\text{x}$ is divided into $N$ bins of fixed width $\Delta \log a$.  Each bin has a characteristic grain size, $a_j$, given by its log-average grain size, and the number density of each bin is calculated using \eqref{eq:mrnn}.  

The mass of a grain particle is given by
\begin{equation}
m_\text{g} = \frac{4}{3}\pi a_\text{g}^3 \rho_\text{b},
\end{equation}
where $\rho_\text{b}$ is the grain bulk density.

The ion and electron number densities calculated by this process vary as (e.g. \citealp{UmebayashiNakano1980}; \citealp{FujiiOkuzumiInutsuka2011})
\begin{eqnarray}
\frac{{\rm d}n_s}{{\rm d}t} &=& \zeta n_\text{n} - k_{\text{e}s}n_s n_{\text{e}_\text{R}} \notag \\ &-& \sum_{Z=-1}^1 \sum_j k_{s\text{g}}(Z,a_j)n_s n_\text{g}(Z,a_j),  \label{eq:dnidt}\\ 
\frac{{\rm d}n_{\text{e}_\text{R}}}{{\rm d}t}&=& \zeta n_\text{n} - \sum_s k_{\text{e}s}n_s n_{\text{e}_\text{R}} \notag \\ &-& \sum_{Z=-1}^1 \sum_j  k_\text{eg}(Z,a_j)n_{\text{e}_\text{R}}n_\text{g}(Z,a_j), \label{eq:dnedt} \\
\frac{{\rm d}n_\text{g}(Z,a_j)}{{\rm d}t}&=& \ -\sum_s \left[k_{s\text{g}}(Z,a_j)n_s n_\text{g}(Z,a_j) \right. \notag \\ &-& \left.k_{s\text{g}}(Z-1,a_j)n_s n_\text{g}(Z-1,a_j) \right] \notag \\ &-& k_{\text{eg}}(Z,a_j)n_\text{e} n_\text{g}(Z,a_j) \notag \\ &+& k_{\text{eg}}(Z+1,a_j)n_\text{e} n_\text{g}(Z+1,a_j)  \label{eq:dngdt}
\end{eqnarray}
where $s \in \left\{ \text{i}_\text{R}, \text{I}_\text{R} \right\}$, $\zeta$ is the ionisation rate, $k_{ij}$ are the charge capture rates and we set $n_\text{g}(Z=\pm2,a_j) \equiv 0$ since their populations will be small compared the singly charged grains.  Although we only include two ion species for simplicity and performance, it is possible to expand this network to include an arbitrary number of molecular ions.

The ion-electron charge capture rates are given by \citep{UmebayashiNakano1990}
\begin{eqnarray}
k_\text{ei}  &=& \left[ 3.5 X \left(\frac{T}{300}\right)^{-0.7} + 4.5 Y \left(\frac{T}{300}\right)^{-0.67}\right]  \notag \\
                   &\times & 10^{-12} \text{cm}^3\text{ s}^{-1} \label{eq:keilight}\\
k_\text{eI} &=& 2.8\times 10^{-12}\left(\frac{T}{300}\right)^{-0.86}  \text{cm}^3\text{ s}^{-1} , \label{eq:keiheavy}
\end{eqnarray}
where $X$ and $Y$ are the mass fractions of hydrogen and helium, respectively, used to determine the mass of the light ion.  For grains, the charge capture rates are \citep{FujiiOkuzumiInutsuka2011}
\begin{eqnarray}
k_{s\text{g}}(Z,a_j)  &=& a_j^2 \sqrt{\frac{8\pi k_\text{B}T}{m_s}}\left\{ \begin{array}{l l} \left(1-\frac{Z e^2}{a_j k_\text{B}T}\right) & \text{if }Z \le 0\\  
                                                              \exp\left(\frac{-Z e^2}{a_jk_\text{B}T}\right) & \text{if }Z > 0  \end{array}\right. ,  \label{eq:kigl}\\
k_\text{eg}(Z,a_j) &=&a_j^2 \sqrt{\frac{8\pi k_\text{B}T}{m_\text{e}}}  \left\{ \begin{array}{l l} \exp\left(\frac{Z e^2}{a_j k_\text{B}T}\right) & \text{if }Z < 0 \\
                                                              \left(1+\frac{Z e^2}{a_j k_\text{B}T}\right) & \text{if }Z \ge  0 \end{array}\right., \label{eq:keg}
\end{eqnarray}
where $k_\text{B}$ is the Boltzmann constant, $e$ is the electron charge, $m_\text{e}$ is the mass of an electron, $Z$ is the charge on the grains and $T$ is the temperature (where the gas and dust are assumed to be in thermal equilibrium).  

From charge neutrality, 
\begin{equation}
\label{eq:chargeneutrality}
\sum_s n_s - n_{\text{e}_\text{R}} + \sum_{Z=-1}^1 \sum_j  Z n_\text{g}(Z,a_j) = 0,
\end{equation}
and from conservation of grains, 
\begin{equation}
\label{eq:graincons}
n_\text{g}(a_j) = \sum_{Z=-1}^1 n_\text{g}(Z,a_j),
\end{equation}
where $n_\text{g}(a_j)$ is the total grain number density as calculated in \eqref{eq:ngrain} or \eqref{eq:mrnn}.

Following \citet{KeithWardle2014}, we assume that the system is approximately in a steady state system (i.e. $\frac{{\rm d}}{{\rm d}t}  \approx 0$).  The set of equations, \eqref{eq:dnidt}-\eqref{eq:dngdt}, \eqref{eq:chargeneutrality} and \eqref{eq:graincons} represents an over-subscription, thus we remove $\frac{{\rm d}n_{\text{e}_\text{R}}}{{\rm d}t}$ and $\frac{{\rm d}n_\text{g}(Z=0,a_j)}{{\rm d}t}$ in favour of charge neutrality and conservation of grains.  The number densities are calculated by iteratively solving
\begin{equation}
\label{eq:system}
\bm{n}^{t+1} = \bm{n}^t - \mathbb{J}^{-1}(\bm{n}^t) \bm{f}(\bm{n}^t) = \bm{n}^t - \bm{x}^t
\end{equation}
where $t$ is the iteration number, $\mathbb{J}(\bm{n}^t)$ is the Jacobian, and in the default case, $\bm{n}^{t} = \left\{ n_{\text{i}_\text{R}}^t, n_{\text{I}_\text{R}}^t,n_\text{e}^t,n_\text{g}^t(Z=-1),n_\text{g}^t(Z=0),n_\text{g}^t(Z=-1)\right\}$  and $\bm{f}(\bm{n}^t) = 0$ are $\left\{ \frac{{\rm d}n_{\text{i}_\text{R}}}{{\rm d}t}\right.$, $\frac{{\rm d}n_{\text{I}_\text{R}}}{{\rm d}t}$, charge neutrality, $\frac{{\rm d}n_\text{g}(Z=-1)}{{\rm d}t}$, grain conservation, $\left.\frac{{\rm d}n_\text{g}(Z=1)}{{\rm d}t}\right\}$.   Rather than solving for inverse of the Jacobian, we use LU decomposition of the Jacobian to solve for $\bm{x}^t$ in $\mathbb{J}(\bm{n}^t) \bm{x}^t = \bm{f}(\bm{n}^t)$.

The number densities can span several orders of magnitude, and this method becomes unstable for $n_\text{g} >> n_\text{e}$.  Although this will likely not be important in most physical applications, an alternative method using average grain charges is described in Appendix \ref{app:Zgrain}; this method is only used after the above method fails to iterate to a solution within a set number of iterations.

%-----------------------------------------------------------------------------------------
\subsection{Thermal ionisation}
\label{sec:algo:IT}
At high gas temperatures, thermal ionisation can remove electrons.  For each species, $j$, the ion number densities can be calculated using the Saha equation, 
\begin{equation}
\label{eq:saha}
\frac{n_{\text{e}_\text{T}} n_{\text{i}_\text{T},j,k+1}}{n_{\text{i}_\text{T},j,k}} = \frac{2g_{j,k+1}}{g_{j,k}} \left( \frac{2\pi m_\text{e} k_\text{B} T}{h^2}\right)^{3/2} \exp{\left( -\frac{\chi_{j,k+1}}{k_\text{B}T} \right)},
\end{equation}
where $n_{\text{i}_\text{T},j,k}$ is the number density of species $j$ which has been ionised $k$ times, $\chi_{j,k}$ is the ionisation potential, $g_{j,k}$ is the statistical weight of level $k$, and $h$ is Planck's constant.  We assume that each atom can be ionised at most twice, thus the total number density of species $j$ will be $n_{\text{i}_\text{T},j} = n_{\text{i}_\text{T},j,0} + n_{\text{i}_\text{T},j,1} + n_{\text{i}_\text{T},j,2}$, and the number of electrons contributed from species $j$ is $n_{\text{e}_\text{T},j} = n_{\text{i}_\text{T},j,1} + 2n_{\text{i}_\text{T},j,2}$.   Since the number density of each ionisation level of each species can be written as a function of electron number density only, the total electron number density from thermal ionisation is
\begin{equation}
n_{\text{e}_\text{T}} = \sum_j \left[n_{\text{i}_\text{T},j,1}\left(n_{\text{e}_\text{T}} \right) + 2n_{\text{i}_\text{T},j,2}\left(n_{\text{e}_\text{T}} \right)\right],
\end{equation}
which can then be solved iteratively using the Newton-Raphson method.  The singly and doubly ionised ions, with number densities $n_{\text{i}_\text{T},1}$ and $n_{\text{i}_\text{T},2}$, respectively, are tracked separately since they have different charges.

The average ion mass is given by  \citep{DraineSutin1987}
\begin{equation}
m_{\text{i}_\text{T}} = \left(\frac{n_{\text{i}_\text{T},1} + n_{\text{i}_\text{T},2}}{\sum_{j,k} \frac{ n_{\text{i}_\text{T},j,k} }{ \sqrt{m_{\text{i}_\text{T},j}} }}\right)^2,
\end{equation}
where $m_{\text{i}_\text{T},j}$ is the mass of a particle of species $j$; note that all ionisation levels are assumed to have the same mass.

In their analytical discussion, \citet{KeithWardle2014} account for the possibility that dust grains absorb electrons.  Similar to cosmic ray ionisation, this would lead to a reduction in the electron number density and a population of negatively charged grains.  However, electron absorption by grains is not currently included in our thermal ionisation model.  At high temperatures, the grains absorb less than a few percent of the electrons, thus do not contribute meaningfully to $n_{\text{e}_\text{T}}$.  At low temperatures, $Z_{\text{i}_\text{T}} = 0$ since $n_{\text{e}_\text{T}} = 0$, but numerical round-off error leads to $Z_{\text{i}_\text{T}} \approx 0$, which then gives non-sensical results of the electron number density.  Given the physical irrelevance and numerical difficulties at high and low temperatures, respectively, electron absorption is not included in this version of \nicil.
%-----------------------------------------------------------------------------------------
\subsection{Charged species populations}
Both cosmic ray and thermal ionisation are calculated independently of one another for numerical stability and efficiency.  This results in two disconnected free electron populations, and the total electron number density is given by $n_\text{e}  = n_{\text{e}_\text{R}} + n_{\text{e}_\text{T}}$.  As will be shown in Section~\ref{sec:tests}, one ionisation process typically dominates the other, thus our method is reasonable.  

The ionisation processes yield seven charged populations:
\begin{enumerate}
\item free electrons with mass $m_\text{e}$, number density $n_\text{e}$ and charge $Z_{\text{e}} \equiv -1$,
\item light ions from cosmic ray ionisation with mass $m_{\text{i}_\text{R}}$, number density $n_{\text{i}_\text{R}}$ and charge $Z_{\text{i}_\text{R}} = 1$,
\item metallic ions from cosmic ray ionisation with mass $m_{\text{I}_\text{R}}$, number density $n_{\text{I}_\text{R}}$ and charge $Z_{\text{I}_\text{R}} = 1$,
\item singly ionised ions from thermal ionisation with mass $m_{\text{i}_\text{T}}$, number density $n_{\text{i}_\text{T},1}$ and charge $Z_{\text{i}_\text{T},1} = 1$,
\item doubly ionised ions from thermal ionisation with mass $m_{\text{i}_\text{T}}$, number density $n_{\text{i}_\text{T},2}$ and charge $Z_{\text{i}_\text{T},2} = 2$,
\item charged grains from cosmic ray ionisation with mass $m_\text{g}$, number density $n_\text{g}(Z=-1)$ and charge $Z_\text{g} = -1$.
\item charged grains from cosmic ray ionisation with mass $m_\text{g}$, number density $n_\text{g}(Z=1)$ and charge $Z_\text{g} = +1$.
\end{enumerate}
These species are defined with subscripts $\text{e},\text{i}_\text{R},\text{I}_\text{R},\text{i}_{\text{T},1},\text{i}_{\text{T},2}$ and $\text{g}$, respectively, where the latter will be used generally for all grain populations.  The density of neutral particles is
\begin{eqnarray}
\rho_\text{n} &=& \rho - \left(\rho_{\text{i}_\text{R}} + \rho_{\text{I}_\text{R}} + \rho_{\text{i}_\text{T}} + \rho_\text{e}\right), \\
                    &=& \rho - \left[n_{\text{i}_\text{R}}m_{\text{i}_\text{R}} + n_{\text{I}_\text{R}}m_{\text{I}_\text{R}}  + \left(n_{\text{i}_\text{T},1} + n_{\text{i}_\text{T},2}\right)m_{\text{i}_\text{T}} + n_\text{e}m_\text{e}\right]. \notag
\end{eqnarray}

%-----------------------------------------------------------------------------------------
\subsection{Conductivities}
\label{sec:algo:con}

The conductivities are calculated under the assumption that the fluid evolves on a timescale longer than the collision timescale of any charged particle.  The collisional frequencies, $\nu$, are empirically calculated rates for charged species.  The electron-ion rate is \citep{PandeyWardle2008}
\begin{equation}
\nu_\text{ei} = 51 \ \text{s}^{-1} \left(\frac{n_\text{e}}{\text{cm}^{-3}}\right)\left(\frac{T}{\text{K}}\right)^{-3/2},
\end{equation}
and the ion-electron rate is $\nu_\text{ie} = \frac{\rho_\text{e}}{\rho_\text{i}}\nu_\text{ei}$.  Note that $\nu_{\text{ei}_\text{R}} = \nu_{\text{eI}_\text{R}} = \nu_{\text{ei}_\text{T},1} = \nu_{\text{ei}_\text{T},2} \equiv \nu_\text{ei}$ because there is no dependence on ion properties, and $\nu_{\text{i}_\text{R}\text{e}} \ne \nu_{\text{i}_\text{T}\text{e}}$.

The plasma-neutral collisional frequency is 
\begin{equation}
\nu_{j\text{n}} = \frac{\left< \sigma v\right>_{j\text{n}}}{m_\text{n}+m_j}\rho_\text{n},
\end{equation}
where $\left< \sigma v\right>_{j\text{n}}$ is the rate coefficient for the momentum transfer by the collision of particle of species $j$ with the neutrals.  For electron-neutral collisions, it is assumed that the neutrals are comprised of hydrogen and helium, since these two elements should dominate any physical system.  The rate coefficient is then given by
\begin{equation}
\left< \sigma v\right>_\text{en} = X_{\text{H}_2} \left< \sigma v\right>_{\text{e-H}_2} + X_\text{H} \left< \sigma v\right>_{\text{e-H}} + Y \left< \sigma v\right>_\text{e-He},
\end{equation}
where  $X_{\text{H}_2}$, $X_{\text{H}}$ and $Y$ are the mass fractions of molecular and atomic hydrogen and helium, respectively, and $X_{\text{H}_2} +  X_{\text{H}} \equiv X$; $X$ and $Y$ are free parameters if {\tt use\_massfrac=.true.}, otherwise they are calculated from the given abundances.  Following \citet{PintoGalli2008}, 
\begin{eqnarray}
\left< \sigma v\right>_{\text{e-H}_2} &=& \left(0.535+0.203\theta-0.163\theta^2+0.050\theta^3\right) \times 10^{-9} \notag \\
                                                        & \times &T^{1/2} \ \text{cm}^3 \ \text{s}^{-1}  \label{eq:sigmaeH2}, \\
\left< \sigma v\right>_{\text{e-H}} &=& \left(2.841+0.093\theta-0.245\theta^2+0.089\theta^3\right) \times 10^{-9} \notag \\
                                                        & \times &T^{1/2} \ \text{cm}^3 \ \text{s}^{-1} , \\
 \left< \sigma v\right>_\text{e-He}     &=& 0.428 \times 10^{-9} T^{1/2} \ \text{cm}^3 \ \text{s}^{-1}, \label{eq:sigmaeHe}
\end{eqnarray}
where $\theta \equiv \log{T}$ for $T$ in K and the drift velocity is assumed to be zero.

The ion-neutral rate is \citep{PintoGalli2008}
\begin{flalign}
\left< \sigma v\right>_\text{in} =& 2.81\times 10^{-9} \ \text{cm}^3 \ \text{s}^{-1}  &\notag \\
 \times& \left|Z_\text{i}\right|^{1/2}\left[  X_{\text{H}_2}\left(\frac{p_{\text{H}_2}}{\text{\AA}^3}\right)^{1/2}\left(\frac{\mu_{\text{i-H}_2}}{m_\text{p}}\right)^{-1/2}  \right. &\notag \\
+ & \left. X_{\text{H}}\left(\frac{p_{\text{H}}}{\text{\AA}^3}\right)^{1/2}\left(\frac{\mu_{\text{i-H}}}{m_\text{p}}\right)^{-1/2}  \right. &\notag \\
+ & \left. Y \left(\frac{p_\text{He}}{\text{\AA}^3}\right)^{1/2}\left(\frac{\mu_\text{i-He}}{m_\text{p}}\right)^{-1/2} \right],&
\end{flalign}
where $p_{\text{H}_2}$, $p_{\text{H}}$ and $p_\text{He}$ are the polarisabilities \citep{Osterbrock1961}.  This term is calculated independently for each ion due to the ion mass dependence in the reduced masses, $\mu_{\text{i-H}_2}$, $\mu_{\text{i-H}}$ and $\mu_\text{i-He}$.

For grain-neutral collisions, the rate coefficient is (\citealp{WardleNg1999}; \citealp{PintoGalli2008})
\begin{equation}
\left< \sigma v\right>_\text{gn} = \pi a_\text{g}^2\delta_\text{gn}\sqrt{\frac{128k_\text{B}T}{9\pi m_\text{n}}},
\end{equation}
where $\delta_\text{gn}$ is the Epstein coefficient, and 
\begin{equation}
m_\text{n} = \left(\frac{X}{m_{\text{H}_2}}+ \frac{Y}{m_\text{He}} + \sum_k \frac{Z_k}{m_k}\right)^{-1}
\end{equation}
is the mass of the neutral particle, where we sum over all included metals, $k$ (if {\tt use\_massfrac=.false.}), which have mass fractions $Z_k$ and masses $m_k$.  This is the only instance in this paper where $Z$ does not refer to charge.  Since $m_\text{n}$ is a characteristic mass, we assume that all the hydrogen is molecular hydrogen, which is reasonable for low temperatures.

The Hall parameter for species $j$ is given by 
\begin{equation}
\beta_j = \frac{|Z_j|eB}{m_j c}\frac{1}{\nu_{j\text{n}}}. \label{eq:beta}
\end{equation}
A modified form is presented in Appendix \ref{app:beta}.

Finally, the Ohmic, Hall and Pedersen conductivities are given by (e.g. \citealp{WardleNg1999}; \citealp{Wardle2007})
\begin{eqnarray}
\sigma_\text{O} &=& \frac{ec}{B}\sum_j n_j|Z_j|\beta_j, \\
\sigma_\text{H} &=& \frac{ec}{B}\sum_j \frac{n_j Z_j}{1+ \beta_j^2}, \\
\sigma_\text{P} &=& \frac{ec}{B}\sum_j \frac{n_j |Z_j| \beta_j}{1+ \beta_j^2}.
\end{eqnarray}
We explicitly note that $\sigma_\text{O}$ and $\sigma_\text{P}$ are positive, whereas $\sigma_\text{H}$ can be positive or negative.  The total conductivity perpendicular to the magnetic field is 
\begin{equation}
\sigma_\bot = \sqrt{\sigma_\text{H}^2 + \sigma_\text{P}^2}.
\end{equation}

%-----------------------------------------------------------------------------------------
\subsection{Non-ideal MHD coefficients}
\label{sec:algo:coef}
The induction equation in magnetohydrodynamics is  
\begin{equation}
\frac{{\rm d} \bm{B}}{\text{d} t} = \left(\bm{B}\cdot\bm{\nabla}\right)\bm{v}-\bm{B}\left(\bm{\nabla}\cdot\bm{v}\right) + \left.\frac{\text{d} \bm{B}}{\text{d} t}\right|_\text{non-ideal}, \label{eq:ind}
\end{equation}
where $ \left.\frac{\text{d} \bm{B}}{\text{d} t}\right|_\text{non-ideal}$ is the non-ideal MHD term, which is the sum of Ohmic resistivity (OR), the Hall effect (HE), and ambipolar diffusion (AD) terms; these terms are given by
\begin{eqnarray}
\label{eq:nonideal}
\left.\frac{\text{d} \bm{B}}{\text{d} t}\right|_\text{OR} &=& -\bm{\nabla} \times \left[  \eta_\text{OR}      \left(\bm{\nabla}\times\bm{B}\right)\right], \label{eq:ohm} \\
\left.\frac{\text{d} \bm{B}}{\text{d} t}\right|_\text{HE} &=& -\bm{\nabla} \times \left[  \eta_\text{HE}       \left(\bm{\nabla}\times\bm{B}\right)\times\bm{\hat{B}}\right],  \label{eq:hall} \\
\left.\frac{\text{d} \bm{B}}{\text{d} t}\right|_\text{AD} &=&   \bm{\nabla} \times \left\{ \eta_\text{AD}\left[\left(\bm{\nabla}\times\bm{B}\right)\times\bm{\hat{B}}\right]\times\bm{\hat{B}}\right\}, \label{eq:ambi}
\end{eqnarray}
where the use of the magnetic unit vector, $\bm{\hat{B}}$, is to ensure that all three coefficients, $\eta$, have units of area per time. 

From conservation of energy, the contribution to internal energy, $u$, from the non-ideal MHD terms \citep{WPA2014} is
\begin{eqnarray}
\label{eq:nonideal}
\left.\frac{\text{d} u}{\text{d} t}\right|_\text{OR} &=& - \frac{\eta_\text{OR}}{\rho}\bm{J}\cdot\bm{J} = - \frac{\eta_\text{OR}}{\rho} J^2, \label{eq:u:ohm} \\
\left.\frac{\text{d} u}{\text{d} t}\right|_\text{HE} &=& - \frac{\eta_\text{HE}}{\rho}\left(\bm{J}\times\hat{\bm{B}}\right)\cdot\bm{J} = 0,  \label{eq:u:hall} \\
\left.\frac{\text{d} u}{\text{d} t}\right|_\text{AD} &=& - \frac{\eta_\text{AD}}{\rho}\left[\left(\bm{J}\times\hat{\bm{B}}\right)\times\hat{\bm{B}}\right]\cdot\bm{J}  \label{eq:u:ambi} \\
                                                                         &=&  - \frac{\eta_\text{AD}}{\rho}\left[\left(\bm{J}\cdot\hat{\bm{B}}\right)^2 - J^2\right] \notag
\end{eqnarray}

Next, we invoke the strong coupling approximation, which allows the medium to be treated using the single fluid approximation.  It is assumed that the ion pressure and momentum are negligible compared to that of the neutrals, i.e. $\rho \sim \rho_\text{n}$ and $\rho_\text{i} \ll \rho$, where $\rho$, $\rho_\text{n}$ and $\rho_\text{i}$ are the total, neutral and ion mass densities, respectively.

The general form of the coefficients \citep{Wardle2007} is
\begin{eqnarray}
\eta_\text{OR} &=& \frac{c^2}{4\pi\sigma_\text{O}}, \label{eq:etaOR} \\
\eta_\text{HE} &=& \frac{c^2}{4\pi\sigma_\bot}\frac{\sigma_\text{H}}{\sigma_\bot}, \label{eq:etaHE} \\
\eta_\text{AD} &=& \frac{c^2}{4\pi\sigma_\bot}\frac{\sigma_\text{P}}{\sigma_\bot} - \eta_\text{OR}  \notag \\
                       &=& \frac{c^2}{4\pi \sigma_\text{O}}\frac{\sigma_\text{O}\sigma_\text{P} - \sigma_\bot^2}{\sigma_\bot^2} \equiv \frac{c^2}{4\pi\sigma_\text{O}}\frac{\sigma_\text{A}}{\sigma_\bot^2}. \label{eq:etaAD}
\end{eqnarray}
The value of $\eta$ depends on the microphysics of the model, and $\eta_\text{HE}$ can be either positive or negative, whereas $\eta_\text{OR}$ and $\eta_\text{AD}$ are positive (e.g. \citealp{WardleNg1999}).  To prevent $\eta_\text{AD} \lesssim 0$ due to numerical round-off when $\sigma_\text{O}\sigma_\text{P} \approx \sigma_\bot^2$, we use
\begin{eqnarray}
\sigma_\text{O}\sigma_\text{P} - \sigma_\bot^2 &\equiv& \sigma_\text{A}  \notag \\
&=& \left(\frac{ec}{B}\right)^2\sum_{j>k}  \left[ \frac{ n_j|Z_j|\beta_j}{1+\beta_j^2}\frac{n_k|Z_k|\beta_k}{1+\beta_k^2}\right. \notag \\
                       &\times& \left.\left(\frac{Z_k\beta_k}{|Z_k|} - \frac{Z_j\beta_j}{|Z_j|}\right)^2\right],
\end{eqnarray}
where all pairs of charged species, $j$ and $k$, are summed over (Wardle, private comm.).

The non-ideal MHD terms constrain the timestep by
\begin{equation}
\text{d}t < C_\text{non-ideal}\frac{h^2}{|\eta|},
\end{equation}
where $h$ is the particle smoothing length or the cell size, $\eta = \max\left(\eta_{\text{OR}},|\eta_{\text{HE}}|,\eta_{\text{AD}}\right)$ and $C_\text{non-ideal} < 1$ is a positive coefficient analogous to the Courant number.  Stability tests by \citet{Bai2014} suggest $C_\text{non-ideal} \lesssim \frac{1}{6}$ is required, while wave calculations and tests by \citet{WPB2016} suggest $C_\text{non-ideal} \lesssim \frac{1}{2\pi}$ is required for stability.  Thus, stability can be achieved by choosing a value of $C_\text{non-ideal}$ appropriate for a conditionally stable algorithm.

%-----------------------------------------------------------------------------------------
\section{TESTS PROGRAMMES AND RESULTS}
\label{sec:tests}
The \nicil \ library includes two stand-alone test codes, which can be used to test the effects of varying the parameters.  Both can be compiled by typing {\tt make} in the NICIL directory.  This yields the executables {\tt nicil\_ex\_eta} and {\tt nicil\_ex\_sph}.  The latter is a simple SPH code that is primarily used to demonstrate how \nicil \ is to be implemented into an existing code. 

The test programme {\tt nicil\_ex\_eta} will calculate the non-ideal MHD coefficients, and output them and the constituent components that are required for their calculation, including number densities of all the charged species and elements, the conductivities and the coefficients.  By default, it outputs three sets of data: one over a range of densities assuming a constant temperature, one over a range of temperatures assuming a constant density, and one where temperature and density are related through the barotropic equation of state \citep{MachidaInutsukaMatsumoto2006}:
\begin{equation}
\label{eq:baro}
T =  T_0\sqrt{1+\left(\frac{n}{n_1}\right)^{2\Gamma_1}}\left(1+\frac{n}{n_2}\right)^{\Gamma_2}\left(1+\frac{n}{n_3}\right)^{\Gamma_3}
\end{equation}
where $T_0 = 10$ K, $n$ is the total density, $n_1 = 10^{11}$, $n_2 = 10^{16}$ and $n_3 = 10^{21}$ cm$^{-3}$, $\Gamma_1 = 0.4$, $\Gamma_2 = -0.3$ and $\Gamma_3 = 0.56667$; the fixed temperature and density for the first two data sets can be modified in {\tt src/}\nicileta.   When a magnetic field is required for the constant density and temperature plots, for the purpose of illustration, we use the relation used in \citet{WardleNg1999}:
\begin{equation}
\label{eq:Btest}
\left(\frac{B}{\text{mG}}\right) = \left\{ \begin{array}{l l} \left(\frac{n_\text{n}}{10^6 \text{ cm}^{-3}}\right)^{1/2}; 	    &  n_\text{n} < 10^6 \text{ cm}^{-3} \\
                                                                                     \left(\frac{n_\text{n}}{10^6 \text{ cm}^{-3}}\right)^{1/4}; 	    &  \text{else} \\
\end{array}\right..
\end{equation}
When a magnetic field is required for the barotropic equation of state, we use \citep{LiKrasnopolskyShang2011}
\begin{equation}
\label{eq:Btest}
\left(\frac{B}{\text{G}}\right) = 1.34 \times 10^{-7} \sqrt{n_\text{n}}.
\end{equation}

\begin{figure*}
\begin{center}
\includegraphics[width=1\columnwidth]{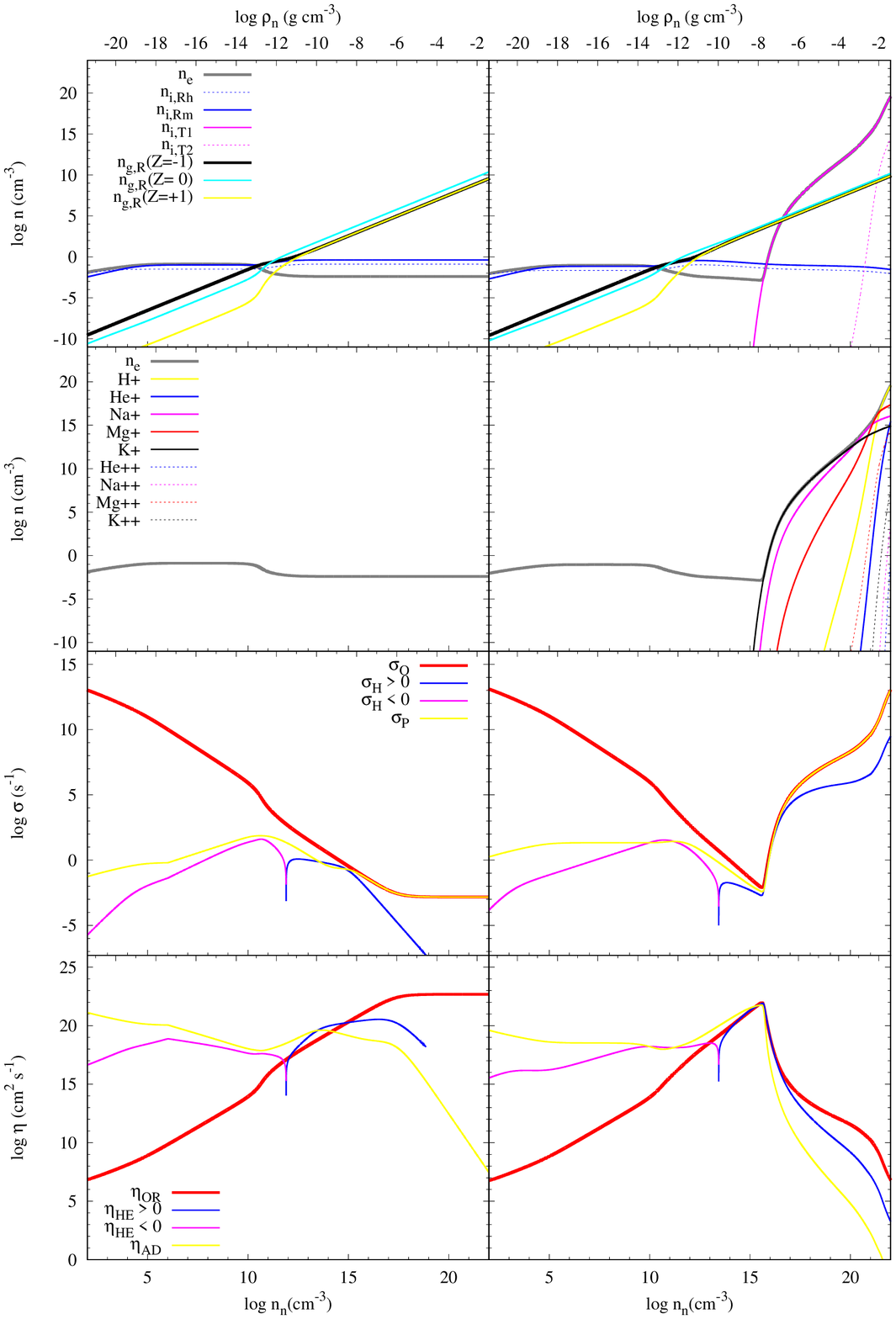}
\includegraphics[width=1\columnwidth]{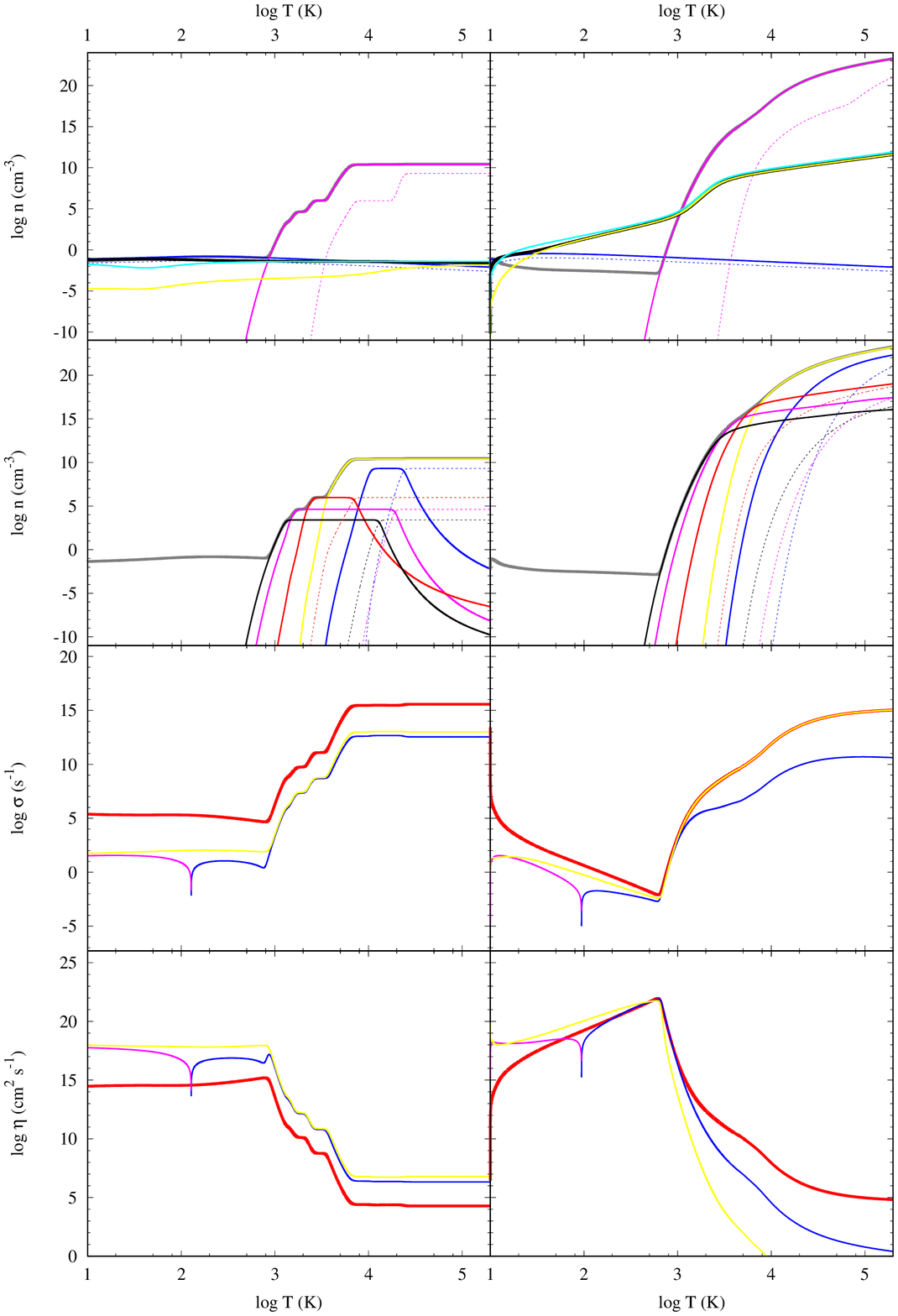}
\caption{\emph{Top to bottom}: Charged species number densities, charged element number densities, conductivities, and non-ideal MHD coefficients.  The first and second columns use $T = 30$ K and $T \equiv T(n)$, respectively, plotted as a function of number density (bottom tics) and mass density (top tics), and the third and fourth columns use $\rho = 10^{-13}$ g cm$^{-3}$ and $n \equiv n(T)$ plotted as a function of temperature.  The second and fourth columns are generated from the same data.  This test was performed using Version 1.2.1 of {\sc Nicil} and the default options listed in Table~\ref{table:app:fp}.}
\label{fig:tests}
\end{center}
\end{figure*}

Fig.~\ref{fig:tests} shows the output of {\tt nicil\_ex\_eta} using the default options listed in Table~\ref{table:app:fp}; the left-hand plot shows the results using $T = 30$ K and $T \equiv T(n)$ as a function of density, and the right-hand plot shows the results using $\rho = 10^{-13}$ g cm$^{-3}$ and $n \equiv n(T)$ as a function of temperature.

For constant $T = 30$ K, the trends for the ion and electron number densities and the conductivities are in approximate agreement with \citet{WardleNg1999}; differences arise as a result of different initial conditions and assumptions.  The elemental number densities are zero, as is expected at such a low temperature.  At this temperature, the conductivities and non-ideal MHD coefficients are strongly dependent on density, while the grain charge and ion and electron number densities essentially have two states: one at high density and one at low density.

For constant $\rho = 10^{-13}$ g cm$^{-3}$,  which is characteristic in discs around protostars, cosmic ray ionisation is the dominant ionisation source for $T \lesssim 600$ K; for  $T \gtrsim 600$ K, thermal ionisation is the dominant source of electrons.  The changeover from the dominance of cosmic ray ionisation to thermal ionisation is abrupt, confirming that the two processes can be calculated independently without loss of information.  

These tests were performed using Version 1.2.1 of \nicil, which can be found in commit 8220ec8 from August 2, 2016.  The plots can be reproduced using the included {\sc Python} graphing script, {\tt plot\_results.py}, which calls {\sc GNUplot}.   It is recommended that the user becomes familiar with the affect of the various parameters and processes prior to implementing \nicil \ into the existing code.  This can be done by modifying the parameters (see Section~\ref{sec:imp:nicil}), plotting them using {\tt plot\_results.py}, and then comparing the new graphs to those made with the default parameters.

%------------------------------------------------------------------------------------------------------------------------------------------------------------------------------------------------------------------------------------------------------------------------------------------------------------
\section{TEST IN A COLLAPSING MOLECULAR CLOUD}
\label{sec:test:mc}

For an astrophysical example, we model the collapse of a rotating, 1M$_\odot$ cloud of gas using the 3D smoothed particle magnetohydrodynamics code {\sc Phantom} with the inclusion of self-gravity.  The cloud has an initial radius of $4\times 10^{16}$ cm and is initially threaded with a magnetic field of strength 163~$\mu$G (5 critical mass-to-flux units) which is anti-aligned with the angular momentum vector.  When the maximum density surpasses $\rho_\text{crit}~=~10^{-10}$~g~cm$^{-3}$ and a given set of criteria are met, the densest gas particle is replaced with a sink particle and its neighbours within $r_\text{acc} =$~6.7~AU are immediately accreted onto it.  This setup is the same as in \citet{WPB2016}, with $10^6$ particles in the cloud.

\begin{figure}
\begin{center}
\includegraphics[width=1.0\columnwidth]{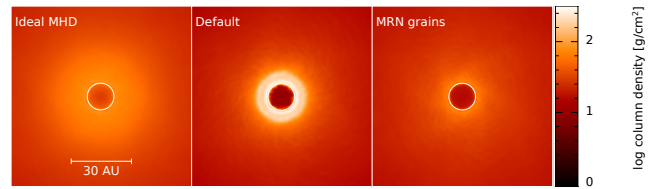}
\caption{Face-on gas column density of a collapsing 1~M$_\odot$ cloud of gas at 1.07$~t_\text{ff}$ (where the free-fall time is $t_\text{ff} = 2.4\times~10^4$~yr).  The magnetic field has a strength of 163$\mu$G (5 critical mass-to-flux units) and is initially anti-aligned with the angular momentum vector.  From left to right, the models use ideal MHD, use the default \nicil \ parameters, and use the MRN grain size distribution with five bins.  The black circle represents the sink particle with the radius of the circle representing the accretion radius of the sink particle.  Each frame is (90~AU)$^2$.  There is no discernible disc in the ideal MHD model, while the model with the default parameters yield the densest disc.}
\label{fig:disc:srho}
\end{center}
\end{figure}

We ran four tests: Ideal MHD, \nicil \ with the default settings, using MRN grain size distribution with five bins, and using only cosmic ray ionisation.  The face-on gas column density at 1.07~$t_\text{ff}$ (where the free-fall time is $t_\text{ff} = 2.4\times~10^4$~yr) is shown in Fig.~\ref{fig:disc:srho} for the ideal, default and MRN models.  No discernible disc forms in the ideal MHD model, which demonstrates the magnetic braking catastrophe (e.g.  \citet{als03}; \citet{PriceBate2007}; \citet{mellonli08}; \citet{hennebellefromang08}; \citealp{WPB2016}).  When the default parameters are used, a weak disc forms around the sink particle.  A low-mass disc may also be forming in the MRN model, however, it is difficult to draw conclusions since such a disc will be heavily influenced by the sink particle.  At this time, the temperature is less than a few hundred Kelvin, thus thermal ionisation plays a minimal role.  As expected, the cosmic ray-only model yields similar results to the default model.

\begin{figure}
\begin{center}
\includegraphics[width=1.0\columnwidth]{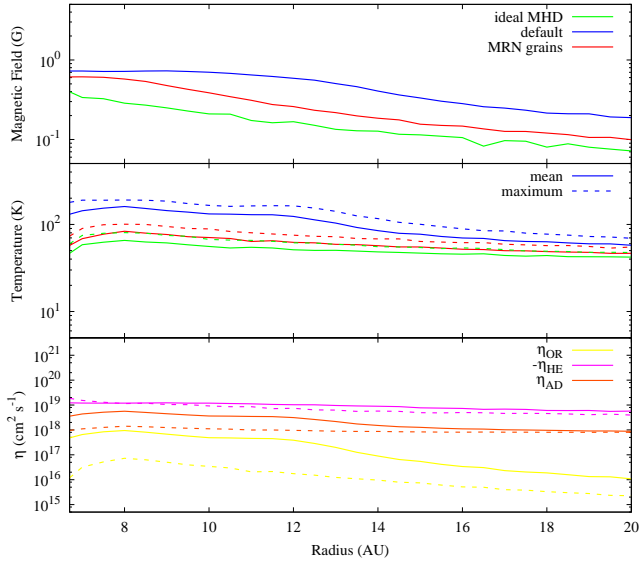}
\caption{Radial profiles of the gas with $\rho > \rho_\text{disc,min}~=~10^{-13}$~g~cm$^{-3}$ at $t =1.07$~$t_\text{ff}$ in the collapsing molecular cloud test.   \emph{Top to bottom}:  Magnetic field, mean and maximum temperature, and non-ideal MHD coefficients for the default (solid) and MRN (dashed) models.  Since $T \lesssim 200$~K, thermal ionisation plays a negligible role.   The coefficients, $\eta$, are larger in the default model, and in both cases, $\eta_\text{HE} < 0$; as a result, the magnetic field strength in the disc increases from the ideal MHD to default to MRN models.}
\label{fig:disc:prof}
\end{center}
\end{figure}

The radial profiles in the disc of the magnetic field, temperature and non-ideal MHD coefficients are shown in Fig.~\ref{fig:disc:prof} for the ideal MHD, default and MRN models.  Gas is assumed to be `in the disc,' if $\rho > \rho_\text{disc,min}~=~10^{-13}$~g~cm$^{-3}$.
Given that the maximum temperature is typically $\lesssim 200$K, thermal ionisation plays a negligible role.  The coefficients are typically larger in magnitude for the default model, with $\left|\eta_\text{HE}\right|  > \eta_\text{AD} > \eta_\text{OR}$.  The negative Hall coefficient is responsible for increasing the magnetic field from the ideal MHD to the MRN to the default model.

This MRN grain model includes five bins, and takes 55 per cent longer to reach the $t = 1.07t_\text{ff}$ than the default model.  Thus, the user must be cautious about performance when using the MRN model.

%------------------------------------------------------------------------------------------------------------------------------------------------------------------------------------------------------------------------------------------------------------------------------------------------------------
\section{IMPLEMENTATION INTO EXISTING CODE}
\label{sec:imp}
\nicil \ is a {\sc Fortran90} module that is contained in one file, {\tt src/}\nicilf.  To include \nicil \ in the user's code, copy this file into the existing code's source directory and include it in the {\tt Makefile}, ensuring that \nicilf \ is compiled in double precision; see {\tt src/Makefile} for details.    

The modifications required to the user's code are summarised below.  The \nicil \ library includes a simple SPH programme, {\tt src/}\nicilsph, which can be used as an example.  

\subsection{Parameter Modifications to \nicilf}
\label{sec:imp:nicil}
The parameters which can be modified by the user are listed at the top of \nicilf \ between the lines labelled `Input Parameters,' and `End of Input Parameters.'  These parameters are defined as {\tt public}, thus can also be modified by other programmes in the user's code.  

If the user wishes to add or remove elements if {\tt use\_massfrac=.false.}, then this is done in the subroutine {\tt nicil\_initial\_species}.  Modify {\tt nelements} to be the new number of elements, and add the characteristics of the new elements to the relevant arrays, using the current elements as a template. 

\subsection{Modifications to user's initialisation subroutine}
\label{sec:imp:init}
\nicil \ contains several variables that require initialising.  When defining the variables and subroutines in the existing initialisation subroutine, include
{\tt  \\ \\ 
use nicil, only : nicil\_initialise \\
integer :: ierr} . \\ \\

\noindent Once the units are initialised by the existing code, include 
{\tt  \\ \\ 
 call nicil\_initialise(utime,udist,umass, \& \\ unit\_Bfield,ierr) \\
if (ierr > 0) call abort\_subroutine\_of\_user \\ \\
}
\noindent where {\tt utime}, {\tt udist}, {\tt umass} and {\tt unit\_Bfield} are unit conversions from CGS to code units.  Note that the code unit of temperature is assumed to be Kelvin.  If {\tt ierr > 0}, then an error occurred during initialisation, and the user's programme should be immediately aborted, using the user's preferred method (e.g. by calling {\tt abort\_subroutine\_of\_user}).

The number densities are calculated iteratively, thus the user must define and save these arrays within the body of their code; the number density array for cosmic rays (thermal ionisation) requires four values (one value) for every cell/particle.  The user is required to initialise these arrays to zero.  For simplicity, they can be treated (i.e. saved and updated) analogously to the user's existing density array.  

\subsection{Modifications to user's density loop}
In many codes, the density is calculated prior to calculating the forces and magnetic fields.  At the same time, \nicil \ should calculate the number densities,  which do not depend on any neighbours.  To calculate the $i^\text{th}$ element of these arrays, {\tt nR} and  {\tt neT},  include
{\tt  \\ \\ 
 call nicil\_get\_ion\_n(rho(i),T(i),\& \\ nR(1:4,i),neT(i),ierr) \\
if (ierr/=0) then \\
\indent call nicil\_translate\_error(ierr) \\
\indent if (ierr > 0) call abort\_subroutine\_of\_user \\
end if \\ \\
}
in the density loop, where {\tt rho(i)} and {\tt T(i)} are the current values of density and temperature, respectively, and {\tt nR(1:4,i)} and {\tt neT(i)} are the characteristic number densities from cosmic rays and the electron number density from thermal ionisation, respectively.  If an error has occurred, {\tt nicil\_get\_ion\_n} will return {\tt ierr}$\ne0$, which will then be passed into {\tt nicil\_translate\_error}.   If {\tt ierr} is returned from {\tt nicil\_translate\_error} as a negative number, then the error is not fatal; the error message will be written to file if {\tt warn\_verbose = .true.}, otherwise all non-fatal errors will be suppressed.  If a fatal error message is returned, then it will be printed to file prior to the user executing the existing code's subroutine to terminate the programme.

The following subroutines and variables need to be imported and defined:
{\tt  \\ \\ 
use nicil, only :  nicil\_get\_ion\_n, \& \\ nicil\_translate\_error\\
integer :: ierr} .

\subsection{Modifications to user's magnetic field loop}
In the subroutine where the magnetic field is calculated, define
{\tt  \\ \\ 
use nicil, only : nicil\_get\_eta, \& \\ nicil\_translate\_error \\
integer:: ierr \\
real:: eta\_ohmi, eta\_halli, eta\_ambii
}\\ \\

Then, for each particle/cell $i$, include 
{\tt  \\ \\ 
call nicil\_get\_eta(eta\_ohmi,eta\_halli,\& \\  eta\_ambii,Bi,rho(i),T(i),nR(1:4,i),neT(i),ierr)\\
if (ierr/=0) then \\
\indent call nicil\_translate\_error(ierr) \\
\indent if (ierr > 0) call abort\_subroutine\_of\_user \\
end if \\  \\
}
where {\tt eta\_ohmi}, {\tt eta\_halli} and {\tt eta\_ambii} are the calculated (i.e. output) non-ideal MHD coefficients, and {\tt Bi} is the magnitude of the particle/cell's magnetic field. 

With the non-ideal MHD coefficients now calculated, the user can calculate  $\left.\frac{\text{d} \bm{B}}{\text{d} t}\right|_\text{non-ideal}$ using the existing routines within the user's code.

\subsection{Optional modifications to the user's code}
\subsubsection{Timesteps}
For each particle/cell, the non-ideal MHD timestep can be calculated by calling
\begin{quotation}
{\tt     call nimhd\_get\_dt(dtohmi,dthalli,dtambii, \& \\ hi,eta\_ohmi,eta\_halli,eta\_ambii) }
\end{quotation}
\noindent where {\tt hi} is the smoothing length of the particle or the size of the cell, and {\tt dtohmi}, {\tt dthalli} and {\tt dtambii} are the three output timesteps.  These are returned independently in case the user employs a timestepping scheme such as super timestepping \citep{AlexiadesEtAl96}, which requires {\tt dtohmi} and {\tt dtambii}  to be treated differently from {\tt dthalli}.
The timesteps can be calculated with minimal extra cost if the modifications are included in the $i$-loop after the non-ideal MHD coefficients are already calculated.

\subsubsection{Internal energy}
\nicil \ also includes a subroutine that evolves the internal energy of the particle or cell.  This is called by
\begin{quotation}
{\tt      call nimhd\_get\_dudt(dudtnonideal, \& \\ eta\_ohmi,eta\_ambii,rhoi,jcurrenti,Bxyzi)} 
\end{quotation}
\noindent where {\tt Bxyzi} is an array containing the three components of the magnetic field, i.e. {\tt (Bxi,Byi,Bzi)}, and {\tt dudtnonideal} is the calculated (i.e. output) change in energy for the particle/cell caused by the non-ideal MHD effects.

\subsubsection{Ion velocity}

The ion velocity is given by
\begin{equation}
\label{eq:vion}
\bm{v}_\text{ion} = \bm{v} + \eta_\text{AD}\left(\bm{\nabla}\times\bm{\hat{B}}\right)\times\bm{\hat{B}},
\end{equation}
which can optionally be calculated by calling
\begin{quotation}
{\tt      call nicil\_get\_vion(eta\_ambii,\& \\vxi,vyi,vzi,Bxi,Byi,Bzi,jcurrenti,\& \\vioni,ierr)} 
\end{quotation}
\noindent where {\tt vxi}, {\tt vyi}, {\tt vzi}, {\tt Bxi}, {\tt Byi}, {\tt Bzi} are the components of the velocity and magnetic field of particle/cell $i$, and {\tt vioni} is an output array containing the three components of the ion velocity.  Similar to {\tt nicil\_get\_dt}, it is optimal to call this array in the $i$-loop after the non-ideal MHD coefficients are calculated for particle $i$.  Note that {\tt ierr} is not initialised in this subroutine, thus must either be passed in from {\tt nicil\_get\_eta}, or reinitialised before calling this routine.  The drift velocity is calculated by
\begin{equation}
\label{eq:vdrift}
\bm{v}_\text{drift} = \bm{v} - \bm{v}_\text{ion}.
\end{equation}

If {\tt warn\_verbose = .true.}, then a warning will be printed if the ion and neutral velocity differ by more than 10 per cent; recall that $v_\text{drift} \approx 0$ is assumed in rate coefficients given in \eqref{eq:sigmaeH2} - \eqref{eq:sigmaeHe}.

\subsubsection{Removing {\tt if}-statements}
Given the versatile design of \nicil, the source code contains several {\tt if}-statements whose argument is a logical operator set prior to runtime.  Since these {\tt if}-statements will be called repeatedly throughout a calculation, their calls may decrease performance of the code if not properly optimised by the compiler.  Thus, {\tt src/hardcode\_ifs.py} can be executed in the {\tt src/} directory to remove them.  

Specifically, this script will read through \nicilsource, determine the value of the logical operators, then rewrite \nicilf \ such that every time an  {\tt if}-statement is encountered with a logical operator, the  {\tt if}-statement will be removed, as will its contents if the logical is set to {\tt .false.}.  Note that any changes made to \nicilf \ will be overwritten, thus all parameters must be set in \nicilsource \  if this script is to be used.  The files \nicilf \ and \nicilsource \ are identical in the repository.

%------------------------------------------------------------------------------------------------------------------------------------------------------------------------------------------------------------------------------------------------------------------------------------------------------------
\section{CONCLUSION}
\label{sec:conc}
We have introduced NICIL: Non-Ideal MHD Coefficients and Ionisation Library.  This a {\sc Fortran90} module that is fully parametrisable to allow the user to determine which processes to include and the values of the parameters.  We have described its algorithms, including the cosmic ray and thermal ionisation processes, and how the conductivities and non-ideal MHD coefficients are calculated.  We have summarised how to implement \nicil \ into an existing code.  

The library contains two test codes, one of which outputs the non-ideal MHD coefficients and its constituent components.  We have described the results using the default values, and showed how the components are affected by temperature and density.  

\nicil \ is a self-contained code that is freely available, and is ready to be implemented in existing codes.  

\begin{acknowledgements}
We would like to thank Daniel Price and Mathew Bate for useful discussions and for their testing and debugging efforts.  JW acknowledges support from the Australian Research Council (ARC) Discovery Projects Grant DP130102078 and from the European Research Council under the European Community's Seventh Framework Programme (FP7/2007- 2013 grant agreement no. 339248).
\end{acknowledgements}
%------------------------------------------------------------------------------------------------------------------------------------------------------------------------------------------------------------------------------------------------------------------------------------------------------------
\begin{appendix}

%-----------------------------------------------------------------------------------------
\section{LIST OF FILES}
\label{app:list}
Table~\ref{table:app:files} lists and summarises the important files that are included in the \nicil \ library.
\begin{table*}
\begin{center}
\begin{tabular}{l l}
\hline\hline
Filename            & Description \\
\hline
{\tt nicil\_ex\_eta}               & Executable: Test programme that will display various properties calculated \\ & with \nicil \\
{\tt nicil\_ex\_sph}              & Executable: Test SPH programme to be used as an example of how to \\ & implement \nicil \ into the user's code \\
\\
{\tt plot\_results.py}           &  {\sc Python} Script: This will allow the user to plot the data obtained from the  \\ & above two executables; this script calls {\sc GNUplot}  \\
{\tt Makefile}                      & Makefile: Secondary makefile that will allow compiling from \nicil's home \\ & directory \\
\\
{\tt src/}\nicilf                     & Source code: The \nicil \ code; this is the file to be copied into the source \\ & directory of the user's code\\
{\tt src/nicil\_souce.F90}    & Source code: Duplicate of \nicil \ code; this will be used to create an updated \\ & \nicil \ if is {\tt src/hardcode\_ifs.py} is run \\
{\tt src/}\nicileta                 & Source code: Test programme that will display various properties calculated \\ & with \nicil \\
{\tt src/}\niciletasup          & Source code: Supplementary subroutines for \nicileta \\
{\tt src/}\nicilsph                & Source code: Test SPH programme to be used as an example of how to \\ & implement \nicil \ into the user's code \\
{\tt src/}\nicilsphsup          & Source code: Supplementary subroutines for \nicilsph \\
{\tt src/Makefile}               & Makefile: Primary makefile that will compile the test codes \\
{\tt src/hardcode\_ifs.py}   & {\sc Python} Script: Creates new \nicilf \ from {\tt nicil\_souce.F90} by removing \\ & {\tt if}-statements of logical operators based upon user's input\\
\\
{\tt Graphs\_Default/}       & Folder: Contains plots using the default parameters \\
{\tt Graphs/}                    & Folder: Empty folder where user's graphs will be saved \\
{\tt data/}                         & Folder: Empty folder where user's output will be written \\
\hline\hline
\end{tabular}
\caption{A list of the important files in the \nicil \ library.  The first two files are executables that appear only after \nicil \ is compiled.}
\label{table:app:files} 
\end{center}
\end{table*}

%-----------------------------------------------------------------------------------------
\section{FREE PARAMETERS}
\label{app:fp}
Table \ref{table:app:fp} lists the free parameters, the default settings, and the reference for the default value (where applicable).
\begin{table*}
\begin{center}
\begin{tabular}{c c l }
\hline\hline
Variable                  & Default value & Description  \\
\hline
{\tt use\_ohm}            & {\tt .true.}       & Calculates and returns $\eta_\text{OR}$ if true  \\
{\tt use\_hall}             & {\tt .true.}       & Calculates and returns $\eta_\text{HE}$ if true \\
{\tt use\_ambi}           & {\tt .true.}       & Calculates and returns $\eta_\text{AD}$ if true  \\
{\tt eta\_constant }     & {\tt .false.}      & Use self-consistently calculated coefficients if false  \\
{\tt g\_cnst}                & {\tt .true.}       & Use constant grain radius if true  \\
{\tt ion\_rays}             & {\tt .true.}       & Calculate ionisation from cosmic rays if true \\
{\tt ion\_thermal}        & {\tt .true.}       & Calculate thermal ionisation if true\\
{\tt use\_massfrac}    & {\tt .false.}      & Use chemical abundances if false  \\
{\tt mod\_beta}          & {\tt .false.}      & Use the modified Hall parameters if true \\
{\tt warn\_verbose}   & {\tt .false.}      & Prints all warnings to file, including when assumptions are violated \\
\\
{\tt fdg}                      & 0.01                                    & $f_\text{dg}$, dust-to-gas mass ratio \\ && \citep{PollackEtAl1994} \\
{\tt a\_grain}              & $10^{-5}$  cm                     & $a_\text{g}$, constant grain radius for {\tt g\_cnst=.true.} \\ && \citep{PollackEtAl1994} \\
{\tt an\_grain}            & $5.0\times 10^{-7}$ cm      & $a_\text{n}$, minimum grain radius for {\tt g\_cnst=.false.} \\ && \citep{WardleNg1999} \\
{\tt ax\_grain}            & $2.5\times 10^{-5}$ cm      & $a_\text{x}$, maximum grain radius for {\tt g\_cnst=.false.}  \\ && \citep{WardleNg1999} \\
{\tt rho\_bulk}            &    3.0  g cm$^{-3}$              & $\rho_\text{b}$, bulk grain density \\ && \citep{PollackEtAl1994} \\
{\tt mass\_ionR\_mp}&  24.3m$_\text{p}$               &  $m_{\text{i}_\text{R}}$, mass of ion for cosmic ray ionisation; \\ && default: mass of magnesium \citep{AsplundEtAl2009} \\
 \\
{\tt zeta\_cgs}           & $10^{-17}$ s$^{-1}$            & $\zeta$, ionisation rate \\ && (c.f. \citealp{KeithWardle2014}) \\
{\tt delta\_gn}            & 1.3                                    & $\delta_\text{g}$, Epstein coefficient for $\left< \sigma v\right>_\text{gn}$ \\ && \citep{LiuEtAl2003} \\
 \\
{\tt massfrac\_X}      &  0.70         & $X$, mass fraction of hydrogen for {\tt use\_massfrac=.true.}  \\
{\tt massfrac\_Y}      &  0.28         & $Y$, mass fraction of helium for {\tt use\_massfrac=.true.}      \\
\\
\hline\hline
\end{tabular}
\caption{A list of the important parameters in \nicil, along with the default values and references.  The first column is the name of the variable in the code, and the actual variable (where it exists) is given in the third column as part of the description.}
\label{table:app:fp} 
\end{center}
\end{table*}

For {\tt use\_massfrac = .false.}, abundances are used in the thermal ionisation calculation.  Five elements are included by default, and they, their abundances and their first and second ionisation potentials are listed in Table~\ref{table:app:abun}.  The abundance, $x_j$ is related to the logarithmic abundance $X_j$ by $x_j = 10^{X_j}/\left(\sum_i  10^{X_i}\right)$.
\begin{table*}
\begin{center}
\begin{tabular}{c c c c c c c c}
\hline\hline
                           &                           & \multicolumn{2}{c}{}                          &  \multicolumn{2}{c}{Ionisation}                       &  \multicolumn{2}{c}{Ratio of }                               \\
                           &                           & \multicolumn{2}{c}{Abundance}        &  \multicolumn{2}{c}{ potential (eV)}                &  \multicolumn{2}{c}{statistical weights}                 \\
Element              & Atomic Number & Logarithmic  & Relative                     & First     & Second                                           & $g_1/g_0$               & $g_2/g_1$                        \\
\hline
Hydrogen, H       &  1                       & 12.00           & $9.21 \times 10^{-1}$ & 13.60  & -                                                      &    1/2                        &        -                                 \\
Helium, He          &  2                      & 10.93           & $7.84 \times 10^{-2}$  & 24.59  & 54.42                                              &    2/1                        &       1/2                              \\
Sodium, Na         & 11                      &  6.24           & $1.60 \times 10^{-6}$  &   5.14  & 47.29                                              &    1/2                        &        6/1                              \\
Magnesium, Mg  & 12                      &  7.60           & $3.67 \times 10^{-5}$  &   7.65  & 15.03                                              &    2/1                        &        1/2                              \\
Potassium, K      & 19                      &  5.03           & $9.87 \times 10^{-8}$  &   4.34  & 31.62                                              &    1/2                        &        6/1                              \\
\hline\hline
\end{tabular}
\caption{The default abundances for {\tt use\_massfrac=.false.} (\citealp{Cox2000}; \citealp{AsplundEtAl2009}; \citealp{KeithWardle2014}).}
\label{table:app:abun} 
\end{center}
\end{table*}

We initially assume that all of the hydrogen is molecular hydrogen, H$_2$.  Using the local temperature and the Saha equation, the number densities of both molecular and atomic hydrogen are determined, viz.,
\begin{equation}
\label{eq:saha}
\frac{n_\text{H} n_\text{H}}{n_{\text{H}_2}} = \left[ \frac{2\pi \left( m_\text{H}m_\text{H}/m_{\text{H}_2}  \right) k_\text{B} T}{h^2}\right]^{3/2} \exp{\left( -\frac{\chi_{H-H_2}}{k_\text{B}T} \right)},
\end{equation}
where $\chi_{H-H_2} = 4.476$ eV is the dissociation potential and $n_\text{H} + n_{\text{H}_2} = A_\text{H}n_0/2$, where $A_\text{H}$ is the abundance of hydrogen as given in Table~\ref{table:app:abun} and $n_0$ is the total number density.  The mass fractions, $X_{\text{H}_2}$ and $X_\text{H}$, are then calculated for use in the rate coefficients.  For thermal ionisation, we only allow molecular hydrogen to be singly ionised, with an ionisation potential of $15.60$ eV.

%-----------------------------------------------------------------------------------------
\section{CONSTANT NON-IDEAL MHD COEFFICIENTS}
\label{app:fixed}

In some scenarios, such as test problems, it may be useful to use semi-constant coefficients.  To implement this, set {\tt eta\_constant=.true.}; the coefficients then become
\begin{eqnarray}
\eta_\text{OR} &=&  C_\text{OR}, \\
\eta_\text{HE} &=&  C_\text{HE}B, \\
\eta_\text{AD} &=&  C_\text{AD}v_\text{A}^2 ,
\end{eqnarray}
where $B$ is the magnitude of the magnetic field and $v_\text{A} = \frac{B}{\sqrt{4\pi \rho}}$ is the Alfv{\'e}n velocity which are self-consistently calculated, as done during tests of ambipolar diffusion in e.g. \citet{MNKW95}, \citet{ChoiKimWiita2009ApJS}, and \citet{WPA2014}.  $C$ is a free parameter to be input by the user.  The default values are given in Table~\ref{table:app:fixedeta}.
\begin{table*}
\begin{center}
\begin{tabular}{c c l }
\hline\hline
Variable                  & Default value & Description  \\
\hline
{\tt C\_OR}                & 0.1 cm$^2$ s$^{-1}$                                           & $\eta_\text{OR} = C_\text{OR}$   \\
{\tt C\_HE}                &-0.5 cm$^2$ s$^{-1}$                                           & $\eta_\text{HE} = C_\text{HE} B$ \\
{\tt C\_AD}                &  0.01 cm$^2$ s$^{-1}$                                        & $\eta_\text{AD} = C_\text{AD} v_\text{A}^2$  \\
\\
{\tt n\_e\_cnst }         & $1.0\times 10^{19}$  cm$^{-3}$                          &$n_\text{e,0}$, constant electron number density \\
{\tt rho\_i\_cnst  }      & $3.8\times 10^{-11}$  g cm$^{-3}$                      &  $\rho_\text{i,0}$, constant density of ionised gas  \\
{\tt rho\_n\_cnst  }      & $3.8\times 10^{-8}$  g cm$^{-3}$                      &  $\rho_\text{n,0}$, constant density of neutral gas  \\
{\tt alpha\_AD}          & 0                                                                          & $\alpha$, power-law exponent   \\
{\tt gamma\_AD}      & $2.6\times 10^{13}$  cm$^3$  s$^{-1}$ g$^{-1}$ & $\gamma_\text{AD}$, collisional coupling coefficient   \\
{\tt hall\_lt\_zero}      &{\tt .false.}                                                             & Sign of the Hall coefficient  (default: $\eta_\text{HE} > 0$)\\
\hline\hline
\end{tabular}
\caption{The free parameters for fixed resistivity coefficients using {\tt eta\_constant=.true.}.  The top three parameters are used if {\tt eta\_const\_calc = .false.} and the bottom four parameters are used if {\tt eta\_const\_calc = .true.}.}
\label{table:app:fixedeta} 
\end{center}
\end{table*}

To calculate the coefficients from fixed parameters that are characteristic of the environment, set ${\tt eta\_const\_calc = .true.}$, and the coefficients become
\begin{eqnarray}
\eta_\text{OR} &=&  \frac{m_\text{e}c^2}{4\pi e^2 n_\text{e,0}}, \\
\eta_\text{HE} &=&  s_\text{H} \frac{c}{4\pi e n_\text{e,0}}B, \\
\eta_\text{AD} &=&  \frac{1}{4\pi \gamma_\text{AD} \rho_\text{i,0}\left(\frac{\rho_\text{n}}{\rho_\text{n,0}}\right)^\alpha}v_\text{A}^2, \label{eq:ambi_cnst_calc}
\end{eqnarray}
where the free parameters are the fixed electron number density $n_\text{e,0}$, the fixed ion density $\rho_\text{i,0}$, the fixed neutral density $\rho_\text{n,0}$, the power-law exponent $\alpha$ ($\alpha = 0$ for molecular cloud densities and $\alpha=0.5$ for low-density regime; c.f. \citealt{MNKW95}), the collisional coupling coefficient for ambipolar diffusion $\gamma_\text{AD}$ and $s_\text{H}= \pm 1$ is the sign of the Hall coefficient.  Invoking the strong coupling approximation, we set $\rho_\text{n} = \rho$ in \eqref{eq:ambi_cnst_calc}.  The default values are given in Table~\ref{table:app:fixedeta}.

Since these forms of the non-ideal MHD coefficients are not self-consistently calculated, it is recommended that {\tt eta\_constant=.true.} be used for testing purposes only.

%-----------------------------------------------------------------------------------------
\section{APPROXIMATING NUMBER DENSITIES USING AVERAGE GRAIN CHARGES}
\label{app:Zgrain}
 
If {\sc NICIL} fails to calculate the number densities from cosmic rays using the Jacobian method describe in Section~\ref{sec:algo:IR}, then the following method is invoked to approximate the number densities.  This method is used with caution since it makes several assumptions.  First, it is assumed that $n_\text{n} \approx n$.  Second, it is assumed that recombination rate is small compared to the other processes, thus $k_{\text{e}s} = 0$.  Third, the grains of each size are treated as a single population with an average grain charge of $\bar{Z} < 0$.  From \eqref{eq:dnidt} and \eqref{eq:dnedt}, these simplifications allow the ion and electron number densities to be written as
\begin{eqnarray}
n_s(\bar{Z}) &=& \frac{\zeta n}{\sum_j k_{s\text{g}}(\bar{Z},a_j) n_\text{g}(a_j)}, \label{eq:altns}\\
n_{\text{e}_\text{R}}(\bar{Z}) &=& \frac{\zeta n}{\sum_j k_\text{eg}(\bar{Z},a_j) n_\text{g}(a_j)}, \label{eq:altne}
\end{eqnarray}
where $s \in \left\{ \text{i}_\text{R}, \text{I}_\text{R} \right\}$.  Then, from charge neutrality,
\begin{equation}
\label{eq:chargeneutrality_app}
\sum_s n_s(\bar{Z})  - n_{\text{e}_\text{R}}(\bar{Z}) + \sum_j \bar{Z}n_\text{g}(a_j) = 0,
\end{equation}
the Newton-Raphson method can be used to solve for $\bar{Z}$.  Recall that $n_\text{g}(a_j)$ is the total grain number density of grains of size $a_j$.

Using $\bar{Z}$, the ion and electron number densities are calculated using \eqref{eq:altns} and \eqref{eq:altne}, respectively.  Using \eqref{eq:dngdt}, charge neutrality and grain conservation, the grain number densities are calculated viz.,
\begin{eqnarray}
n_\text{g}(Z=-1,a_j) &=& \frac{k_\text{eg}^+n_\text{e}(1+\bar{Z})n_\text{g}}{ \sum_s k_{s\text{g}}^-n_s +  k_\text{eg}^-n_\text{e} + 2k_\text{eg}^0n_\text{e}} \\
n_\text{g}(Z=+1,a_j) &=& \frac{(1-\bar{Z})n_\text{g}\sum_s k_{s\text{g}}^0 n_s }{ \sum_s \left[k_{s\text{g}}^+ +2k_{s\text{g}}^0\right]n_s + k_\text{eg}^+ n_\text{e} } \\
n_\text{g}(Z=0,a_j) &=& n_\text{g} - n_\text{g}(Z=-1,a_j) \notag \\ &-& n_\text{g}(Z=1,a_j),
\end{eqnarray}
where $k^+\equiv k(Z=1,a_j)$, $k^-\equiv k(Z=-1,a_j)$ and $k^0\equiv k(Z=0,a_j)$.

%-----------------------------------------------------------------------------------------
\section{MODIFIED HALL PARAMETER}
\label{app:beta}

A modified version of the Hall parameter was used in \citet{WPB2016}:
\begin{subequations}
\label{eq:betamod}
\begin{eqnarray}
\beta_\text{e} &=& \frac{|Z_\text{e}|eB}{m_\text{e} c}\frac{1}{\nu_\text{en}+\nu_\text{ei}}, \label{eq:betae} \\
\beta_\text{i} &=& \frac{|Z_\text{i}|eB}{m_\text{i} c}\frac{1}{\nu_\text{in}+\nu_\text{ie}}, \label{eq:betai} 
\end{eqnarray}
\end{subequations}
With these modifications, the coefficient for Ohmic resistivity from \citet{PandeyWardle2008} and \citet{KeithWardle2014} can be recovered under the assumption $\beta_\text{i} \ll \beta_\text{e}$.  By default, these forms are not used, but are included for legacy reasons.

\end{appendix}
%------------------------------------------------------------------------------------------------------------------------------------------------------------------------------------------------------------------------------------------------------------------------------------------------------------
\bibliographystyle{apj}
\bibliography{Wbib.bib}

\end{document}